\documentclass[prd,aps,nofootinbib,showpacs,11pt]{revtex4}
\usepackage{amsmath,amssymb,graphicx,epsfig}

\usepackage{graphics}
\usepackage{rotating}
\usepackage{epsfig}

\begin{document}

\title{ Generalized Heavy-to-Light Form Factors in Light-Cone Sum Rules}
\author{ 
Ulf-G. Mei{\ss}ner$^{a,b}$,  Wei Wang$^{a}$  
}
\affiliation{ 
  $^a$ Helmholtz-Institut f\"ur Strahlen- und Kernphysik and Bethe Center for
Theoretical Physics, Universit\"at Bonn, D-53115 Bonn, Germany\\
$^b$ Institute for Advanced Simulation, Institut f\"ur Kernphysik and J\"ulich
Center for Hadron Physics, Forschungszentrum J\"ulich,
D-52425 J\"ulich, Germany
}

\begin{abstract}
We study the form factors for a heavy meson into the S-wave  
$  K\pi/\pi\pi$  system with an invariant mass below 1~GeV.  The mesonic final state interactions are described in terms of the scalar form factors, which are obtained from unitarized chiral perturbation theory.  
Employing generalized light-cone distribution amplitudes, we compute the heavy-to-light transition using light-cone sum rules. Our  approach  simultaneously respects  constraints from analyticity and unitarity, and also takes advantage of  the  power expansion in the $1/m_b$ and the strong coupling constant. 
\end{abstract}

\maketitle

\noindent {\it Introduction} -- $B$ decays  into a light vector meson are of particular  interest as they  can  provide valuable information to extract  the Standard Model (SM) parameters and therefore  test the SM. In the case that large  deviations from the SM calculations are found, these will shed light on  new physics scenarios.  Examples for such type of decays include e.g.
the process  $B\to \rho (\to \pi\pi)l\bar\nu$ for the extraction of the CKM matrix element $|V_{ub}|$,  the  reaction $B\to K^*(\to K\pi)l^+l^-$  to test the chirality structure in weak interaction,   and the decay  $B_s\to J/\psi\phi(\to K\bar K)$  to determine  the $B_s-\bar B_s$ mixing phase.   
Recent experimental data on   these channels can be found in  Refs.~\cite{delAmoSanchez:2010af,Sibidanov:2013rkk,Aaij:2013iag,Aaij:2013oba}.

Due to the short lifetime, the light  vector meson can not be directly detected by experiments and  must  be reconstructed from the two or three pseudo-scalars $\pi/K$ final state. Thus  these  decay modes  are at least  four-body processes and the semi-leptonic ones are refereed to as $B_{l4}$ decays in the literature~\cite{Lee:1992ih} (for a recent dispersion theoretical approach
to this reaction, see Ref.~\cite{Kang:2013jaa}). To select candidate events and suppress the combinatorial background, experimentalists often implement  kinematic cuts  on the invariant mass. During this procedure  various partial waves of the $K\pi/\pi\pi$  system may get entangled and bring dilutions to physical observables. 
Particularly it is very likely  the S-wave contributions are of great  importance~\cite{Stone:2008ak,Stone:2009hd,Xie:2009fs,Colangelo:2010bg,Colangelo:2010wg,Leitner:2010fq,Fleischer:2011au,Liu:2013nea,Lu:2011jm,Doring:2013wka,Meissner:2013pba,Li:2010ra,Becirevic:2012dp,Matias:2012qz,Blake:2012mb,Bobeth:2012vn,Jager:2012uw,Descotes-Genon:2013vna,Faller:2013dwa,Zhang:2013iga,Bediaga:2004bc,Magalhaes:2011sh,Kamano:2011ih}. Therefore it is mandatory to have   reliable and accurate predictions considering the high precision  achieved or to be achieved  by experiments.

Decay amplitudes for semi-leptonic $B$ decays into two light-pseudoscalar mesons show two distinctive features. 
On the one hand, the final state interaction of the two  pseudo-scalars should  satisfy  unitarity and analyticity. On the other hand, the  $b$ mass scale is much higher than the hadronic scale, which allows an expansion of  the hard-scattering kernels in terms of the strong coupling constant and the dimensionless power-scaling  parameter $\Lambda_{\rm QCD}/m_b$.  
In this paper, we aim  to develop a formalism that makes use of both these advantages. It simultaneously  combines the perturbation theory at the $m_b$ scale based on the operator product expansion  and the low-energy effective theory inspired by  the chiral symmetry  to describe the S-wave $\pi\pi$ and $K\pi$ scattering.  
For concreteness,  we will choose the $B\to K\pi$ matrix elements  with the $K\pi$ invariant mass below 1~GeV  as an example in the following, while other processes including the charm meson decay can be treated in an analogous way. 
If the factorisation can be proved,  these form factors will also play an important role in the study of charmless three-body $B$ decays~\cite{Chen:2002th,ElBennich:2009da,Bediaga:2013ela,Cheng:2013dua}.

\medskip
  
\noindent{\it Generalized  form factor} -- The  matrix elements 
\begin{eqnarray}
    \langle (K\pi)_0(p_{K\pi})|\bar s \gamma_\mu\gamma_5 b|\overline B  (p_B)
 \rangle  &=& -i  \frac{1}{m_{K\pi}} \bigg\{ \bigg[P_{\mu}
 -\frac{m_{B }^2-m_{K\pi}^2}{q^2} q_\mu \bigg] {\cal F}_{1}^{B\to K\pi}(m_{K\pi}^2, q^2) \nonumber\\
 &&
 +\frac{m_{B }^2-m_{K\pi}^2}{q^2} q_\mu  {\cal F}_{0}^{B\to K\pi}(m_{K\pi}^2, q^2)  \bigg\},\nonumber\\ 
  \langle (K\pi)_0(p_{K\pi})|\bar s \sigma_{\mu\nu} q^\nu \gamma_5 b|\overline B  (p_B)
 \rangle  &=& -   \frac{{\cal F}_T^{B\to K\pi}(m_{K\pi}^2, q^2)}{m_{K\pi}(m_B+m_{K\pi})}  \big[q^2 P_{\mu}- (m_{B }^2-m_{K\pi}^2) q_\mu \big],
 \label{eq:generalized_form_factors}
\end{eqnarray}
define the S-wave generalized form factors ${\cal F}_i$~\cite{Doring:2013wka}. 
Here,  $P=p_B+ p_{K\pi}$ and $q=p_B- p_{K\pi}$.   

The $K\pi$ system  with invariant mass below 1 GeV can be treated as a light hadron  and  more explicitly in the  kinematics region we are considering,  the $m_{K\pi}$ is small and the $K\pi$ system moves very fast, the soft-collinear effective theory (SCET) is applicable~\cite{Bauer:2000ew,Bauer:2000yr,Bauer:2001yt,Beneke:2003pa}.  As shown later this $K\pi$ system has similar  light-cone distribution amplitudes  with the ones for  a light hadron.  The transition matrix elements for $B\to K\pi$ may   be factorized in the same way as the ordinary  $B$-to-light ones like the $B\to \pi$ transition. 
It has been demonstrated in   SCET that, in the soft contribution limit, the form factors obey   factorization~\cite{Beneke:2000wa,Bauer:2002aj,Beneke:2003pa}: 
\begin{eqnarray}
F_i = C_i \xi(q^2) + \Delta F_i, 
\end{eqnarray}
where $C_i$ are the short-distance and calculable functions, and $\xi$ is a universal soft form factor from the large recoil symmetry in the heavy quark $m_b\to\infty$  and large energy  $E\to\infty$ limit~\cite{Charles:1998dr}. Symmetry breaking terms, starting at order $\alpha_s$,  can be encoded into $\Delta F_i$, and  can be expressed as a convolution  in terms of the LCDA~\cite{Beneke:2000wa,Bauer:2002aj,Beneke:2003pa,Beneke:2004rc,Beneke:2005gs}. 



Watson's theorem implies that  phases measured in the $K\pi$ elastic scattering and in a decay channel where the
$K\pi$ system decouple with other hadrons  are equal (modulo
$\pi$ radians). This leads to
\begin{eqnarray}
 \langle (K\pi)_0 |\bar s \Gamma b|\overline B\rangle   
 \propto F_{K\pi}(m_{K\pi}^2),
\end{eqnarray}
where  the strangeness-changing scalar form factors are defined by
\begin{eqnarray}
 \langle 0| \bar sd |K\pi\rangle = C_X B_0 F_{K\pi}(m_{K\pi}^2)~. 
\label{defff}
\end{eqnarray} 
$C_X$ is an isospin factor and $B_0$ is proportional to the QCD condensate parameter.  For  the $K^- \pi^+$,   $C_X=1$. 
Below the $K+3\pi$ threshold, about 911 MeV, the $K \pi$ scattering  is strictly elastic. The inelastic contributions in the $K \pi$ scattering comes from the $K+3\pi$ or $K\eta$.  In the region from 911 MeV to 1 GeV, the $K+3\pi$ channel has a limited phase space, and thus is generically suppressed.    Moreover, as a process-dependent study, it has been demonstrated the states with two additional pions will  not give sizeable    contributions to physical observables~\cite{Bar:2012ce}. 
Though differences may be expected, some similarities  might be shared. We leave the $K+3\pi$ contributions for future work. The $K\eta$ coupled-channel effects can be included in the unitarized approach of chiral perturbation theory~\cite{Jamin:2000wn,Jamin:2001zq,Jamin:2006tj,Bernard:2007tk,Bernard:2009ds}.

In the following we will choose the light-cone sum rules (LCSR) to calculate the  ${\cal F}_i$. An analysis  in  other approaches like  the $k_T$ factorisation~\cite{Keum:2000ph,Keum:2000wi,Lu:2000em,Lu:2000hj,Kurimoto:2001zj} would be similar,  and for recent developments in this approach see Refs.~\cite{Nandi:2007qx,Li:2010nn,Li:2012nk,Wang:2010ni,Wang:2012ab,Li:2012md,Wang:2013ix,Kim:2013ria,Li:2013xna}. 
As a reconciliation of the original QCD sum rule approach~\cite{Shifman:1978bx,Shifman:1978by}  and the
application of perturbation theory to hard processes,  LCSR
exhibit  several advantages in the calculation of quantities like  the meson form 
factors~\cite{Craigie:1982ng,Braun:1988qv,Chernyak:1990ag,Belyaev:1994zk,Colangelo:2000dp}.   In the hard scattering region the operator product expansion (OPE) near
the light-cone is applicable. Based on the light-cone OPE, form 
factors are expressed as a convolution of  light-cone distribution amplitudes (LCDA) with a perturbatively  calculable hard kernel. The leading twist  and a few sub-leading twist LCDA give the dominant contribution, while higher twist terms are suppressed.

The calculation begins with the correlation function: 
\begin{eqnarray}
 \Pi(p_{K\pi},q)&=& i \int d^4x \, e^{iq\cdot x}    \langle ({K\pi})_0(p_{K\pi})|  {\rm
 T}\left\{j_{\Gamma_1}(x),j_{\Gamma_2}(0)\right\}|0\rangle, 
 \label{corr}
\end{eqnarray}
where $j_{\Gamma_1}$ is one of the currents  
in Eq.~\eqref{eq:generalized_form_factors}  defining
the form factors: $j_{\Gamma_1}=\bar s\gamma_\mu\gamma_5b$
for ${\cal F}_1$ and ${\cal F}_0$, and $j_{\Gamma_1}=\bar
s\sigma_{\mu\nu}\gamma_5 q^\nu b$ for ${\cal F}_T$.   We choose
$j_{\Gamma_2}=\bar b i\gamma_5 d$ to interpolate the $B$ meson, whose  matrix element  gives  the decay 
constant $f_{B}$:
\begin{eqnarray}
 \langle \overline B(p_{B})| \bar b i\gamma_5 d|0\rangle &=&
 \frac{m_{B}^2}{m_{b}+m_d}f_{B}~.
\end{eqnarray}

The hadronic representation of the correlation function consists in  the contribution of the $B$ meson and of the
higher resonances and the continuum  state: 
\begin{eqnarray}
  \Pi^{\rm HAD}(p_{K\pi},q) &=&   \frac{\langle {(K\pi)_0}(p_{K\pi})|j_{\Gamma_1}|\overline
 {B}( p_{K\pi}+q)\rangle \langle \overline {B}(p_{K\pi}+q)|j_{\Gamma_2}|0\rangle}
 {m_{B}^2-(p_{K\pi}+q)^2}\nonumber \\
 &&+ \int_{s_0}^\infty ds \frac{\rho^h(s,q^2)}
 {s-(p_{K\pi}+q)^2}, \label{hadronic}
\end{eqnarray}
where  higher resonances and  the continuum of states are
described in terms of the spectral function $\rho^h(s,q^2)$ and start from the threshold $s_0$.  
 
The correlation function  in Eq.~\eqref{corr} can also be
evaluated in the deep Euclidean region in QCD  at the quark level.  
The quark-hadron duality guarantees the equality of the two calculations and thus we obtain the  sum rules
\begin{eqnarray}
  &&  {\langle {(K\pi)_0}(p_{K\pi})|j_{\Gamma_1}|\overline {B}(p_{B})\rangle \langle \overline {B}(p_{B})|j_{\Gamma_2}|0\rangle}
 {\rm
 exp}\left[-\frac{m_{B}^2}{M^2}\right]\nonumber \\
 &=&\frac{1}{\pi}\int_{(m_b+m_s)^2}^{s_0}ds \,
 {\rm exp}[-s/M^2] \, \, {\rm Im}\Pi^{\rm QCD}(s,q^2)~.  \label{SR-generic}
\end{eqnarray}
In the above, a Borel transformation has been performed to  improve  the convergence of the OPE series, 
and to  enhance  the contribution of the low-lying states to the correlation function for suitably chosen values of $M^2$.

The calculation of $\Pi^{\rm QCD}$ is based on the expansion of the
T-product in the correlation function  near the light-cone, which  produces
matrix elements of non-local quark-gluon operators. 
These quantities  are in terms of the generalized LCDA  of
increasing twist~\cite{Diehl:1998dk,Polyakov:1998ze,Kivel:1999sd,Diehl:2003ny}: 
\begin{eqnarray}
 \langle {(K\pi)_0}|\bar s (x)\gamma_\mu d(0)|0\rangle &=&  N
 p_{{K\pi}\mu} \frac{1}{m_{K\pi}}
 \int_0^1 due^{i up_{K\pi}\cdot x}\Phi_{K\pi}(u),
 \nonumber\\
 \langle {(K\pi)_0}|\bar s (x)  d(0)|0\rangle &=&N  
 \int_0^1 du e^{iup_{K\pi}\cdot x}\Phi_{K\pi}^s(u),
 \nonumber \\
 \langle {(K\pi)_0}|\bar s(x)\sigma_{\mu\nu}   d(0)|0\rangle &=&
 - N\frac{1}{6}  (p_{{K\pi}\mu} x_\nu -p_{{K\pi}\nu} x_\mu) \int_0^1 du e^{i up_{K\pi}\cdot x} {\Phi_{K\pi}^\sigma(u)} ,
\end{eqnarray}
where  $N=C_X B_0F_{K\pi}$.  
Due to the Watson's theorem, the above  matrix elements are proportional to the $K\pi$ scalar form factors which have been absorbed into the normalisation constant $N$.  As a result, the distribution amplitudes,  $\Phi_{K\pi}$ and $\Phi_{K\pi}^{s,\sigma}$, are real. 

The LCDA $\Phi_{K\pi}$ is twist-2, and the other two  are twist-3. Their normalisations are given as 
\begin{eqnarray}
 \int_0^1 du \Phi_{K\pi}(u)= \frac{m_{s}-m_{d}}{m_{K\pi}},\nonumber\\
 \int_0^1 du \Phi_{K\pi}^s(u)=\int_0^1
 du\Phi_{K\pi}^\sigma(u)=1.
\end{eqnarray}
The  use of conformal symmetry in QCD~\cite{Braun:2003rp} indicates  that the twist-3 LCDA  have the asymptotic form~\cite{Diehl:1998dk,Polyakov:1998ze,Kivel:1999sd,Diehl:2003ny}:
\begin{eqnarray}
 \Phi_{K\pi}^s(u)=1, \nonumber\\ \Phi_{K\pi}^\sigma(u)= 6u(1-u),
\end{eqnarray}
and the twist-2 LCDA can be expanded in terms of Gegenbauer moments:
\begin{eqnarray}
 \Phi_{K\pi}(u)= 6u(1-u) \sum_{n} a_n C_n^{3/2}(2u-1)~.
\end{eqnarray}
It is  worthwhile  to stress that these generalized LCDA for a two-hadron system  have the same form as the ones for a light meson~\cite{Diehl:1998dk,Polyakov:1998ze,Kivel:1999sd,Diehl:2003ny}. 

\medskip
  
\noindent{\it Results} -- For the sake of presentation, we define 
\begin{eqnarray}
 {\cal F}_{i}(q^2, m_{K\pi}^2) = C_XB_0 m_{K\pi} F_{K\pi}(m_{K\pi}^2)  \overline   F_i (m_{K\pi}^2,q^2),
\end{eqnarray}
with the expressions 
\begin{widetext}
\begin{eqnarray}
 \overline   F_+&=& N_F
  \bigg\{\int_{u_0}^1\frac{du}{u}{\rm exp}\left[-\frac{m_b^2+u\bar u m_{K\pi}^2-\bar uq^2}{uM^2}\right]  \bigg[-m_b\Phi_{K\pi}(u)+um_{K\pi}\Phi_{K\pi}^s(u)+\frac{1}{3}m_{K\pi}\Phi_{K\pi}^\sigma(u) \nonumber\\
  &&   +\frac{
 m_b^2+q^2-u^2m_{K\pi}^2}{uM^2}\frac{m_{K\pi}\Phi_{K\pi}^\sigma(u)}{6}
 \bigg] 
 +\exp{[-s_0/M^2]}\frac{m_{K\pi}\Phi_{K\pi}^\sigma(u_0)}{6}\frac{m_b^2-u_0^2m_{K\pi}^2+q^2}
 {m_b^2+u_0^2m_{K\pi}^2-q^2}\bigg\},
 \label{eq:fplus} 
 \end{eqnarray}
 \end{widetext}

\begin{widetext}
\begin{eqnarray}
 \overline  F_-&=& N_F\left\{\int_{u_0}^1\frac{du}{u}{\rm
 exp}\left[-\frac{m_b^2+u\bar u m_{K\pi}^2-\bar uq^2}{uM^2}\right]
 \bigg[ m_b\Phi_{K\pi}(u)+(2-u) m_{K\pi}\Phi_{K\pi}^s(u)\right. \nonumber\\
 &&\;\;\;\left. +\frac{1-u}{3u}m_{K\pi}\Phi_{K\pi}^\sigma(u) -\frac{u({m_b^2+q^2-u^2m_{K\pi}^2})+2(
 m_b^2-q^2+u^2m_{K\pi}^2)}{u^2M^2}\frac{m_{K\pi}\Phi_{K\pi}^\sigma(u)}{6}
 \bigg]\right.\nonumber\\
 &&\left. -\frac{ u_0({m_b^2+q^2-u_0^2m_{f_0}^2})+2(
 m_b^2-q^2+u_0^2m_{K\pi}^2) }{u_0(m_b^2+u_0^2m_{K\pi}^2-q^2)}
  \exp{[-s_0/M^2]}\frac{m_{K\pi}\Phi_{K\pi}^\sigma(u_0)}{6}\right\},
  \label{eq:fminus}
  \\
 \overline  F_T&=&2 N_F (m_{B}+m_{K\pi}) 
 \bigg\{\int_{u_0}^1\frac{du}{u} {\rm exp}\left[-\frac{(m_b^2-\bar uq^2+u\bar
 um_{K\pi}^2)}{uM^2}\right]
 \left[-\frac{\Phi_{K\pi}(u)}{2}+m_b\frac{m_{K\pi}\Phi_{K\pi}^\sigma(u)}{6uM^2}\right] \nonumber\\
 &&   +m_b\frac{m_{K\pi}\Phi_{K\pi}^\sigma(u_0)}{6}
   \frac{\exp[-s_0/M^2]}{m_b^2-q^2+u_0^2m_{K\pi}^2}\bigg\},
   \label{eq:ftensor}
\end{eqnarray}\end{widetext}
where
\begin{eqnarray}
 N_F &=&\frac{m_b+m_s}{2m_{B}^2f_{B}} {\rm
 exp}\left[\frac{m_{B}^2}{M^2}\right],\nonumber\\
 u_0&=&\frac{m_{K\pi}^2+q^2-s_0+\sqrt{(m_{K\pi}^2+q^2-s_0)^2+4m_{K\pi}^2(m_b^2-q^2)}}{2m_{K\pi}^2}~.
 \label{eq:u0}
\end{eqnarray}
Our formulae can be compared to the results for the $B$ to a scalar $\bar qq$ meson transition.   
Quantities including the invariant mass and LCDA for the $K\pi$ system will be replaced by those for the scalar $\bar qq$ resonance as in Ref.~\cite{Wang:2008da,Colangelo:2010bg}.  

\begin{figure}[ht]
\begin{center}
\includegraphics[scale=0.6]{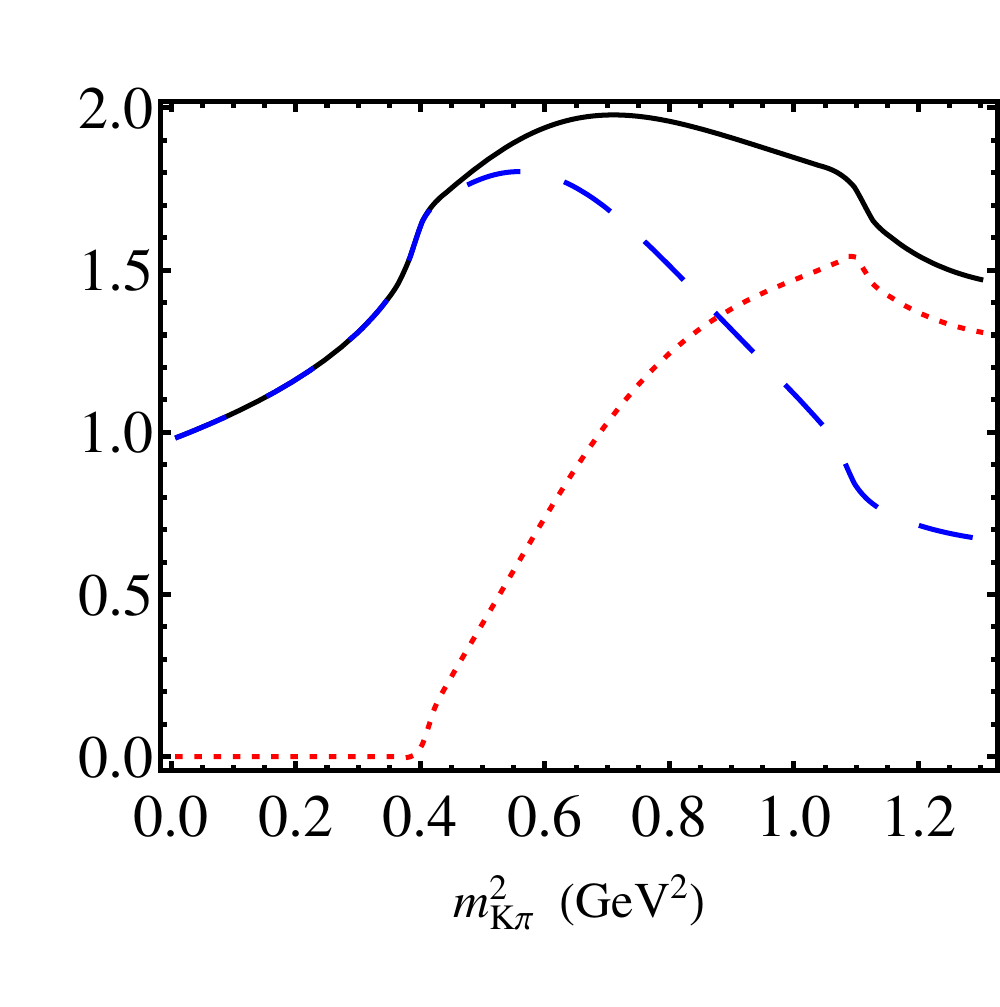}  
\caption{Scalar $K\pi$ form factors calculated in unitarized chiral perturbation theory. Solid, dashed and dotted lines correspond to the magnitude, the real and the imaginary part, in order.} 
\label{fig:scalarFF}
\end{center}
\end{figure}

The   scalar form factor $F_{K\pi}$ has been calculated in the  unitarized approach embedded in the chiral perturbation theory,  and we refer the reader to Ref.~\cite{Doring:2013wka} for details.  We quote  these results displayed  in Fig.~\ref{fig:scalarFF}, where  the solid, dashed and dotted lines correspond to the magnitude, the real and the imaginary part of $F_{K\pi}$, respectively.
From this figure, we can see the imaginary part shows an approximate  linear dependence  on $m_{K\pi}^2$. Such behaviour  can be derived from the calculation in  chiral perturbation theory and we quote the next-to-leading order results~\cite{Gasser:1984ux}:
\begin{eqnarray}
F^\chi_{K\pi }(s)&=& 
1+\frac{4L_5^r\,s}{f^2}+\frac{s}{4\Delta_{K\pi}}
\left(5\mu_\pi-2\mu_K-3\mu_{\eta_8}\right)
+\bar J_{K\pi}K_{K\pi,K\pi}-\frac{1}{3}\,
\bar J_{K\eta_8}K_{K\eta_8,K\pi}  \ ,
\label{ffchi}
\end{eqnarray}
where  $L_5^r$ is a low energy constant,  and 
\begin{eqnarray}
 K_{K\pi ,K\pi} &=& -\frac{1}{8 f^2}\left(2\Sigma-5s
+\frac{3\Delta_{K\pi}^2}{s}\right),\quad 
K_{K\eta_8, K\pi}=-\frac{1}{8 f^2}\left(3s-2\Sigma
-\frac{\Delta_{K\pi}^2}{s}\right)  \ , 
\nonumber\\
  \mu_i &=& \frac{M_i^2}{32\pi^2 f^2} \log\left(\frac{M_i^2}{\mu^2}\right)  \ , 
  \nonumber\\ 
\bar J&=&\frac{1}{32\pi^2}\bigg[2+\left(\frac{M_1^2-M_2^2}{s}
-\frac{M_1^2+M_2^2}{M_1^2-M_2^2}\right)\log\frac{M_2^2}{M_1^2} \nonumber\\
&& 
-\frac{\lambda(s)}{s}\bigg(\log(s+\lambda(s)+M_1^2-M_2^2)
+\log(s+\lambda(s)-M_1^2+M_2^2)
\nonumber\\
&&-\log(-s+\lambda(s)-M_1^2+M_2^2)-\log(-s+\lambda(s)+M_1^2-M_2^2)\bigg)\bigg],
\end{eqnarray}
and $\Sigma=M_\pi^2+M_K^2$, $\Delta_{K\pi}=M_K^2-M_\pi^2$. 
 $f$ is   the pion decay constant,  $f = 92.4$~MeV,
$\lambda^2(s)=[s-(M_1+M_2)^2][s-(M_1+M_2)^2]$, and $s\equiv s+i\epsilon$ ensures
that the correct sheet of the logarithm is set.
The imaginary part of the  scalar form factor $F^\chi_{K\pi }$  arises from the function  $\bar J$: 
\begin{eqnarray}
 {\rm Im}[\bar J]= \frac{1}{16\pi} \frac{\lambda(s)}{s},
\end{eqnarray}
which   leads to an approximate  linear dependence on the  $m_{K\pi}^2$ below $1{\rm GeV}^2$. 
However, this linear dependence disappears in the region with large $m_{K\pi}^2$ since higher-order contributions become important and are taken into account in the unitarized approach. This  has been  discussed in detail    in Ref.~\cite{Doring:2013wka}.

The $B$ meson decay constant is taken from the Lattice QCD calculation of ref.~\cite{Neil:2011ku}: 
$f_B=(196.9\pm 8.9)$~MeV. 
As demonstrated above, one of the most key inputs is  the two-hadron LCDA.  We will use   asymptotic forms for the twist-3 ones, but no knowledge on the twist-2 is  available at present. In Ref.~\cite{Cheng:2005nb}, the authors have studied the LCDA for the light scalar mesons below 1 GeV in the $\bar qq$ scenario. We shall  use these results in our numerical calculation,  bearing in mind large uncertainties that may be introduced by this approximation.  To the best of our knowledge, there is no available results  on non-asymptotic twist-3 LCDA for scalar mesons below 1 GeV. Studies of LCDA for scalar mesons above 1 GeV can be found in Refs.~\cite{Lu:2006fr,Zhang:2011db,Han:2013zg}, but these results  are not applicable here due to the  large differences in the (invariant) mass.  In the future, we hope the situation can be improved  using nonperturbative QCD tools including    Lattice QCD simulations. 

It is interesting to notice that ${\cal F}_i$ can also be evaluated with other interpolating currents. One example is the chiral current~\cite{Sun:2010nv,Han:2013zg}, which has the advantage of isolating different contributions by twists. In this framework, choosing the suitable current, one can completely smear out the uncertain twist-2 LCDA in the QCD calculation, with the price of the complex hadronic representation since the parity partner of the $B$ meson also contributes to the same correlation function. 

\begin{figure}[ht]
\begin{center}
\includegraphics[scale=0.75]{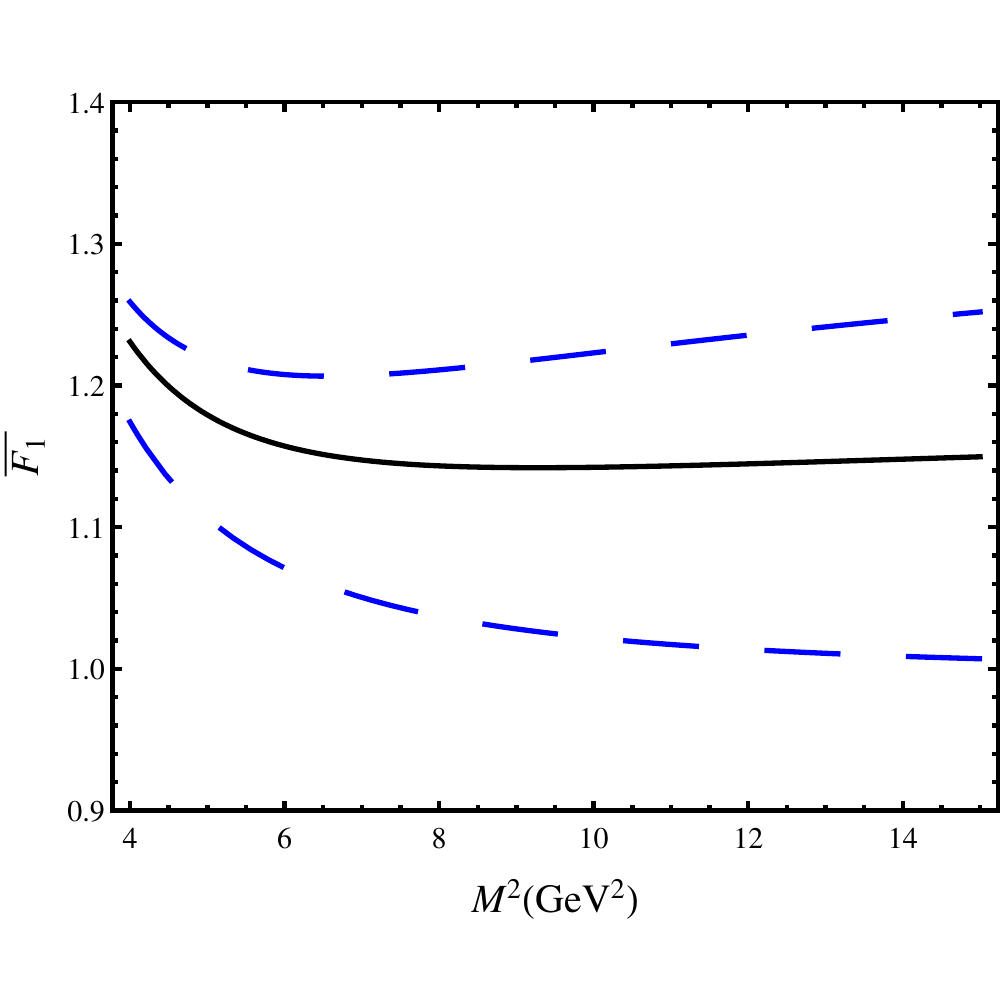}   \hspace{1.cm}
\includegraphics[scale=0.75]{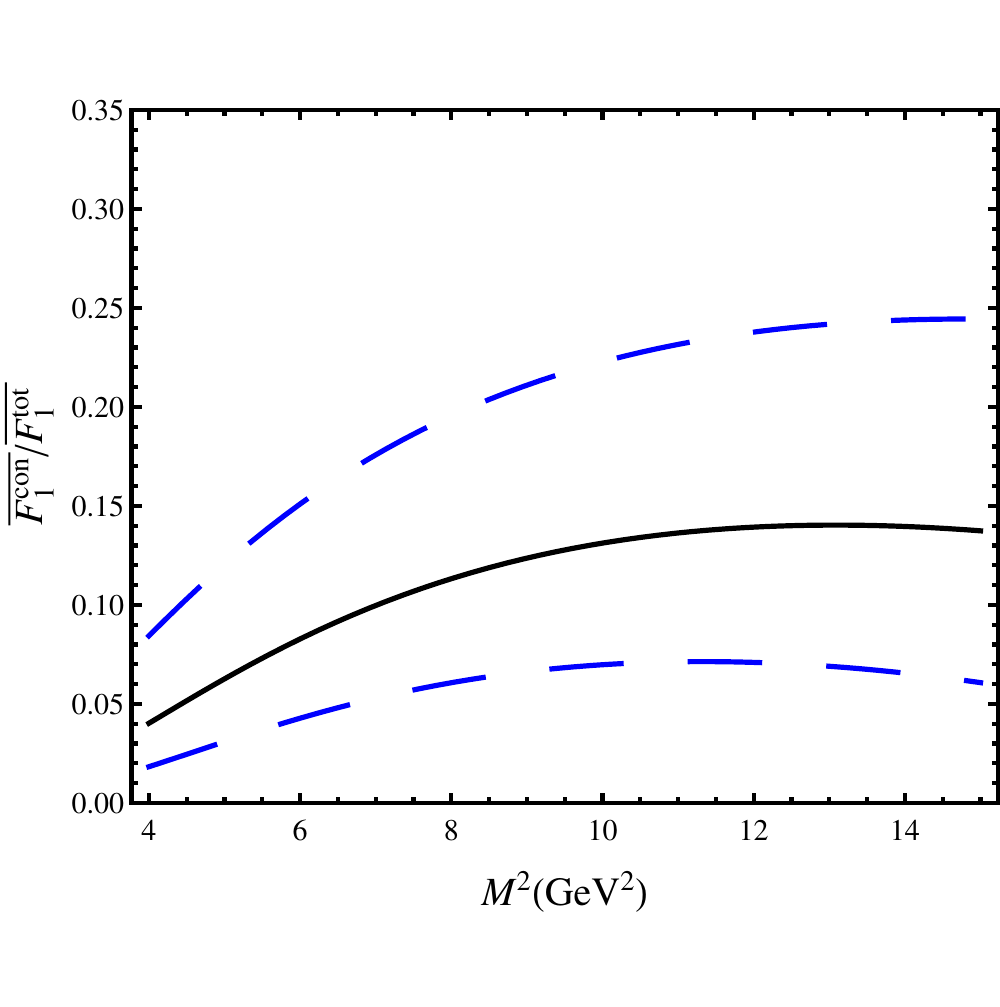}  
\caption{The dependence of the form factor $\bar F_1$ (left panel) and the ratio of the continuum and total contributions (right panel) on the Borel parameter. Solid lines denote the central value while the dashed curves correspond to variations of threshold parameter: $s_0= (34\pm 2) {\rm GeV}^2$.  Results for $\bar F_1$  are   stable when 
$M^2>  6\,  {\rm GeV}^2$, while the continuum contribution is typically  smaller than  $30\%$.  } 
\label{fig:BorelDepen}
\end{center}
\end{figure}

The criteria in LCSR  to find sets of parameters $M^2$ (the Borel
parameter) and $s_0$ (the continuum threshold) is  that the resulting form factor does not
depend   much on the precise values of these parameters;  additionally both  the continuum contribution,
the dispersive integral from $s_0$ to $\infty$ in Eq.~\eqref{hadronic}, and the higher power corrections, arising from the neglected  higher twist  LCDA,  should not be  significant.
One more requirement on the $s_0$ is that it should not be  too much away
from the ``reasonable" value: $s_0$ is to separate the ground state  from higher mass contributions,
and thus  should be below the next known   resonance, in this case, $B_1$ with $J^P=1^+$. Thus approximately this parameter should be close to $33~{\rm GeV}^2$~\cite{Beringer:1900zz}. 
Studies  of ordinary heavy-to-light  form factors in LCSR, see for instance Ref.~\cite{Ball:2004rg},  also  suggested a similar result, ranging from $33~{\rm GeV}^2$ to $36~{\rm GeV}^2$, while some bigger values are derived  in the recent update of $B\to \pi$ form factor in LCSR~\cite{Khodjamirian:2011ub}.

Numerical results  based on LCSR for the auxiliary function  $\overline  F_1$ at the  $K\pi$ threshold $m_{K\pi}= m_{K}+m_{\pi}$  are given in 
Fig.~\ref{fig:BorelDepen}, where the dependence of the form factor $\bar F_1$ (left panel) and the continuum/total ratio (right panel) on the Borel parameter are shown. The continuum contribution to the form factors is  obtained by invoking the quark-hadron duality above the threshold $s_0$ and calculating the correlation function on the QCD side.  Solid lines denote the central value while the dashed curves correspond to variations of threshold parameter: $s_0= (34\pm 2) {\rm GeV}^2$. From this figure, we can see that results for $\bar F_1$  are   stable against the variation of $M^2$ when $M^2> 6 \, {\rm GeV}^2$, and meanwhile the continuum contribution is typically  smaller than  $30\%$.   Unfortunately, due to the lack of knowledge on the 3-particle twist-3  and higher twist  generalized LCDA, we are unable to estimate the power corrections due to these LCDA, and we hope this situation  can be improved with more dedicated studies in the  future.

\begin{figure}[ht]
\begin{center}
\includegraphics[scale=0.7]{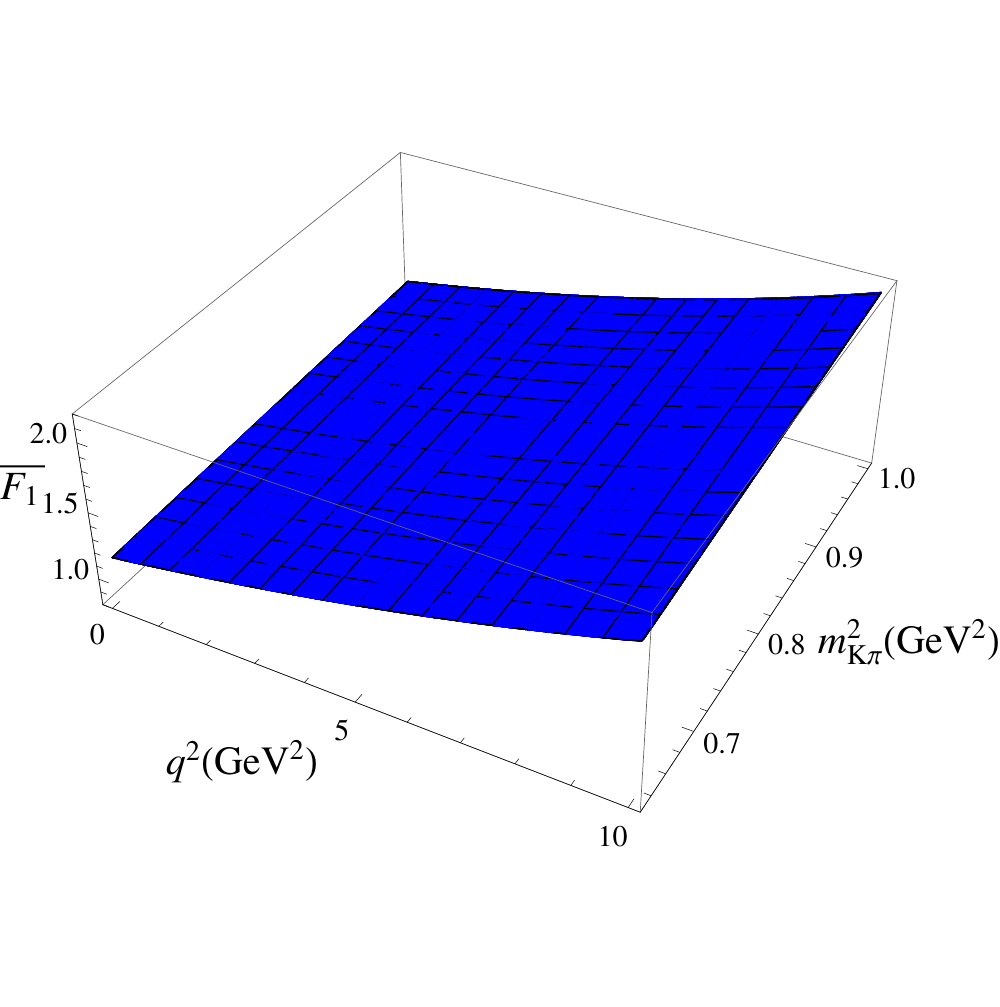} 
\caption{The dependence of $\bar F_1$ on the squared momentum transfer $q^2$ and the two-hadron invariant mass square $m_{K\pi}^2$.} 
\label{fig:MKpiQ2}
\end{center}
\end{figure}

Choosing the value $M^2=8 {\rm GeV}^2$, we show the results in Fig.~\ref{fig:MKpiQ2} for the dependence on the squared invariant mass of the $K\pi$ system and the squared momentum transfer $q^2$. As we can see, the results
 increase with the $q^2$. This behaviour is similar to the $B\to\pi$~\cite{Khodjamirian:2011ub} and  $B\to \rho$~\cite{Ball:2004rg} form factors.    More results and phenomenological   consequences  will be published elsewhere.

\medskip
\noindent{\it Conclusions} -- We have formulated  an  approach to explore  the S-wave generalized form factors for the heavy meson transitions into the  $\pi\pi, K\pi$ final state.    We have adopted unitarized  chiral perturbation theory to account for  the final state  interactions, and include these effects  in the scalar form factors and generalized light-cone distribution amplitudes. The heavy-to-light transition is calculated  within QCD sum rules on the light-cone. Our  approach  simultaneously respects constraints from unitarity and analyticity, and also takes advantage of  the  power expansion in the $1/m_b$ and the strong coupling constant.   With these form factors at hand based on  improved results on the generalized LCDA, one may reliably  explore  the S-wave effects in semi-leptonic heavy meson decays and further non-leptonic charmless three-body $B$ processes if the factorization holds.  
  
\medskip  

\noindent{\it Acknowledgements} -- We thank Michael D\"oring, Feng-Kun Guo, Bastian Kubis and Yu-Ming Wang for  useful discussions. This work is supported in
part by the DFG and the NSFC through funds provided to the Sino-German CRC 110
``Symmetries and the Emergence of Structure in QCD'', the ``EU I3HP Study of
Strongly Interacting Matter''  under the Seventh Framework Program of the EU.

\bigskip
  

\end{document}